\documentclass[prl,twocolumn,groupedaddress,showpacs]{revtex4-1}

\usepackage[latin9]{inputenc}
\usepackage{amsmath}
\usepackage{amssymb}
\usepackage{graphicx}
\usepackage{bm}
\usepackage{color}
\usepackage[hidelinks,colorlinks=true,linkcolor=blue,citecolor=blue,anchorcolor=black,filecolor=cyan,menucolor=red,runcolor=cyan,urlcolor=magenta]{hyperref}

\usepackage[T1]{fontenc}
\usepackage{mathptmx}

\begin{document}
\title{Interaction-enhanced integer quantum Hall effect in disordered systems}

\author{Jun-Hui Zheng}
\author{Tao Qin}
\author{Walter Hofstetter}

\affiliation{Institut f{\"u}r Theoretische Physik, Goethe-Universit{\"a}t, 60438
Frankfurt/Main, Germany.}

\date{\today}

\begin{abstract}
We study transport properties and topological phase transition in two-dimensional interacting disordered systems. Within dynamical mean-field theory, we derive the Hall conductance, which is quantized and serves as a topological invariant for insulators, even when the energy gap is closed by localized states. In the spinful Harper-Hofstadter-Hatsugai model, in the trivial insulator regime, we find that the repulsive on-site interaction can assist weak disorder to induce the integer quantum Hall effect, while in the topologically non-trivial regime, it impedes Anderson localization. Generally, the interaction broadens the regime of the topological phase in the disordered system.  
\end{abstract}

\pacs{xx}

\maketitle

The quantum Hall effect (QHE) in the presence of interaction and disorder has been of great interest for a long time. Interactions play an essential role in the fractional QHE \cite{Tsui1982prl} and disorder is responsible for the existence of the plateaux in the Hall conductance \cite{Huckestein1995rmp,Lee1985rmp,Kramer1993rpp,Arapov2015sc}.  For different models, the perfect quantization of conductance can be violated \cite{Wu2006prl,Xu2006prb,Hohenadler2013jpcm,Neupert2011prl,Johannesson2010prl,Budich2012prb,Geissler2014prb,Nunner2008prl,zheng2016arxiv,Sheng1997prl,Zhang2013cpb} or conversely induced  \cite{Wang2012epl, Kumar2016prb,Raghu2008prl, Martin2016pra,Torma2016prl,Li2009prl,Groth2009prl, Nakai2014prb, Wu2016cpb, Orth2016srs,Chen2017prb,Xie2015prl,Shindou2016prl,Titum2015prl,Titum2016prx,Shen2011prl} 
by disorder and interaction, respectively. Topological invariants are constructed to classify the resulting transport properties \cite{Chiu2016rmp,Wen1990prb,Thouless1982prl} in systems with bulk energy gaps.  General expressions for the invariants of interacting and disordered systems were developed from the perspective of the many-body wave functions (MBW) \cite{Niu1985prb,Sheng2006prl,Lee2008prl,Loring2010epl}. Nonetheless, the MBW can be captured numerically  only for a rather small size of the interacting system.  Equivalent expressions in terms of the single-particle Green's function were developed thereafter,  based on the microscopic theory \cite{Ishkawa1987npb,Wang2010prl, Essin2015}, which
are numerically accessible even for infinite systems if  translational symmetry (TS)  is assumed \cite{Georges1996}.  

Disorder destroys TS, which increases the difficulty of studying topological phase transitions  in  interacting disordered systems. A non-perturbative way of dealing with topological invariants in systems of this type is lacking. In this letter, we focus on the combined effects of  on-site interaction and disorder on a topologically trivial or non-trivial insulator, respectively, by employing dynamical mean-field theory (DMFT). For general systems in the absence of  TS, we derive the Hall conductance within DMFT, where the vertex corrections of the current operator are negligible \cite{Georges1996}. We name the resulting formula {\it generalized Ishikawa-Matsuyama formula} (GIMF). In the presence of a bulk energy gap, the GIMF is quantized and serves as a  topological invariant. Furthermore, as we will prove, localized states do not contribute to the Hall conductance. Thus the GIMF is still a topological invariant even for systems with gapless bulk spectrum, if all states at the Fermi energy are localized.

Numerically, we treat the finite-size disordered system as a supercell of an infinite system \cite{Ceresoli2007prb,Zhang2012prb}. As an example,  we calculate the Hall conductance of the spinful Harper-Hofstadter-Hatsugai model (HHHM) on a square lattice with half filling \cite{Hatsugai1990prb}. By tuning the strength of the on-site staggered potential, the system can be initially prepared in a topologically trivial or non-trivial state. Then the on-site interaction and disorder are added to investigate their effects. Interaction effects are taken into account within DMFT. We find that, for a trivial insulator with weak disorder, the repulsive on-site interaction can assist the disorder to smoothen the staggered potential more efficiently, and thus induce the integer QHE. This result is consistent with the effective medium theory (EMT) we develop,  in which interaction effects are included within the Born approximation. For a topologically non-trivial insulator, the interaction impedes Anderson localization. In general, the interaction broadens the regime of the topological phase in the disordered system. 

In the following we first derive the Hall conductance for systems without TS. The results are derived on a square lattice but can similarly be generalized to any periodic lattice structure.  The Hamiltonian  is 
\begin{equation}
\hat{H}= \hat{H}_0 + \hat{H}_{\rm{int}},
\end{equation}
where $\hat{H}_0 = \sum_{\bm{i}\alpha,\bm{j}\gamma}\hat{c}_{\bm{i}\alpha}^{\dagger}[H_0]_{\bm{i}\alpha,\bm{j}\gamma}\hat{c}_{\bm{j}\gamma}$  is the noninteracting part and $\hat{H}_{\rm{int}}=U\sum_{\bm{i}}\hat n_{\bm{i}\uparrow}\hat n_{\bm{i}\downarrow}$ is the on-site interaction.  $\hat{H}_0=\hat{h}_0 + \hat{H}_{\rm{dis}}$,  where $ \hat{h}_0 = \sum_{\bm{i}\alpha,\bm{j}\gamma}\hat{c}_{\bm{i}\alpha}^{\dagger}[h_0]_{\bm{i}\alpha,\bm{j}\gamma}\hat{c}_{\bm{j}\gamma}$ is the translationally invariant part  and $  \hat{H}_{\rm{dis}}$ represents  disorder.  $\bm{i}$ and $\bm{j}$ are the lattice indices. $\alpha$ and $\gamma$ are spin indices.   $\hat{H}_{0}$ is not diagonal in  momentum space. We denote its matrix elements as $H_{0}(\bm{k}_{1}\alpha,\bm{k}_{2}\gamma) \equiv [H_0]_{\bm{k}_{1}\alpha,\bm{k}_{2}\gamma}$.

Current operators (CO) are derived from the continuity equation and the equation of motion of the local density operator \cite{Bernevig2013book}. The final expression of CO is
\begin{equation}\hat{J}^\mu_{\bm{q}}=\sum_{\bm{k}_1\alpha,\bm{k}_2\gamma}\hat{c}_{(\bm{k}_1+\bm{q}/2)\alpha}^{\dagger}[{j}^{\mu}]_{\bm{k}_1\alpha,\bm{k}_2\gamma} \hat{c}_{(\bm{k}_2-{\bm{q}}/{2})\gamma},
\end{equation}
for $\bm{q}\rightarrow \bm{0}$, where $[j^{\mu}]_{\bm{k}_1\alpha,\bm{k}_2\gamma}=\frac{1}{\sqrt{N}}[\bar\partial_{k_\mu}H_{0}]_{\bm{k}_1\alpha,\bm{k}_2\gamma}$  \cite{jzhengxx}. Here,  $ [\bar\partial_{k_\mu}A]_{\bm{k}_{1}\alpha,\bm{k}_{2}\gamma}\equiv(\partial_{k_{1\mu}}+\partial_{k_{2\mu}}){A}(\bm{k}_{1}\alpha,\bm{k}_{2}\gamma)$ for any matrix $A$ and $\mu = x, y$. $N$ is the total number  of lattice sites. We set $\hbar = e = a =1$, where $\hbar$ is the Planck constant, $a$ is the lattice constant, and $e$ is the electron charge. The CO is the sum of the first derivatives of  $H_0$ with respect to the two momenta. The on-site disorder potential  does not contribute to the CO \cite{jzhengxx}, but disordered tunneling terms do. For the system with TS, the Hamiltonian becomes diagonal in momentum space and the CO recovers  its ordinary form. 

The dc Hall conductance is $\sigma_{H}= -\partial_{\omega}{\rm{Re}}\, \Pi_{yx}(i\omega)|_{\omega = 0^+}$, where $\Pi_{yx}(i\omega_{n})=\lim_{\bm{q}\rightarrow {\bm 0}} \int_{0}^{\beta} d\tau e^{i\omega_{n}\tau} \langle T_{\tau}{\hat{J}_{\bm{q}}^{y}}(\tau){{\hat{J}_{-\bm{q}}^{x}}(0)} \rangle$ 
is the current-current correlation function  \cite{jzhengxx}. The conductance is obtained from the Kubo formula \cite{Mahan2000book} and by using the L\textsc{\char13}H\^opital's rule and the Cauchy-Riemann equations.
Within DMFT, the contribution from the vertex corrections of the CO to the Hall conductance is suppressed by the dimensionality of the system \cite{Georges1996,Kohno1988}. Thus,
$\Pi_{yx} =\frac{1}{\beta}\sum_{n^{\prime}}{\rm Tr} [ j^{y}{G}(i\omega_{n^{\prime}}+i\omega_{n})j^{x}{G}(i\omega_{n^{\prime}})]$, where the trace is  for both  momentum and spin degree of freedom. In real space, the Green's function contains  the on-site self-energy  $\Sigma_{\bm{i}\alpha,\bm{j}\gamma}(i\omega)=\Sigma_{\bm{i}\alpha,\bm{i}\gamma}(i\omega)\delta_{\bm{i} \bm{j}}$ \cite{Hofstetter2008,Georges1996}. In momentum space, we find $[\bar\partial_{k_\mu}\Sigma]_{\bm{k}_1\alpha,\bm{k}_2\gamma} =0$, and thus, $j^{\mu}=\frac{1}{\sqrt{N}}\bar\partial_{k_\mu}{G}^{-1}_{i\omega}$.  Finally, the Hall conductance in the zero temperature limit becomes $\sigma_{H} = {\chi}/{2\pi}$, where 
\begin{equation}
{\chi}\equiv \frac{\epsilon^{\tilde{\mu}\tilde{\nu}\tilde{\rho}}}{6N}\int d\omega {\rm Tr}[{G}_{i\omega}(\bar\partial_{\tilde{\mu}}{G}_{i\omega}^{-1}){G}_{i\omega}(\bar\partial_{\tilde{\nu}}{G}_{i\omega}^{-1}){G}_{i\omega}(\bar\partial_{\tilde{\rho}}{G}_{i\omega}^{-1})]. \label{ChernNum111}
\end{equation}
The indices $\tilde{\mu}$, $\tilde{\nu}$, and $\tilde{\rho}$ run through $k_x$, $k_y$, and $\omega$, and   $ [\bar\partial_{\omega}{G}_{i\omega}]_{\bm{k}_{1}\alpha,\bm{k}_{2}\gamma}\equiv\partial_{\omega}{G_{i\omega}}(\bm{k}_{1}\alpha,\bm{k}_{2}\gamma) $ \cite{jzhengxx}.  We denote the quantity $\chi$ as GIMF.  The GIMF reduces to the original formula \cite{Ishkawa1987npb} when the TS is recovered.

An important result we would like to stress here is that  {\it{the GIMF is quantized and serves as a topological invariant for insulators, regardless whether the bulk spectrum is gapped or gapless.}}  The result is evident for gapped cases  \cite{jzhengxx,Qi2008prb}. For disordered systems, the gap can be closed by disorder. The system can still be an insulator, but with all of states at the Fermi energy being localized. Later we will prove that none of the localized states  contributes to the conductance. Thus, the Hall conductance is still quantized for this specific gapless case \cite{jzhengxx}. 

The above Hall conductivity is derived in the thermodynamic limit. In the numerical approach, the sample for the disorder configuration can be realized only for a finite size system (FSS) with $n_x \times n_y$ lattices.  To apply Eq.\,\eqref{ChernNum111} to the disordered system, we construct an extended infinite system (EIS),  by periodically repeating the FSS in space \cite{Ceresoli2007prb}. The EIS is a quasi-disordered system. The real disordered system can be approached when $n_x$ and $n_y$ become large. The idea is equivalent to introducing twisted boundary conditions to the FSS. In the EIS, the system again becomes periodic in the real space, in which the FSS is a supercell \cite{Ceresoli2007prb,Zhang2012prb}. In the first Brillouin zone, the Bloch wavevector is $\bm{\theta} \equiv (\theta_x,\theta_y)$ with $\theta_\mu \in [0, 2\pi/n_\mu)$ for $\mu = x, y$. Each supercell has $2 n_x  n_y$ internal degrees of freedom, where the factor $2$ is contributed by the spin.  The position $\bm{i}$ in the EIS can be  expressed as $\bm{i}=\bm{R}+\bm{r}$, where $\bm{R}$ is the position of the corresponding supercell and $\bm{r}$ is the relative position in the supercell.  

We define the twisted matrix $A^{\bm\theta}$ as follows,
\begin{equation}
A^{\bm\theta}_{ \bm{r}\alpha, \bm{r}^\prime\gamma}\equiv \sum_{\bm{R}} A_{ (\bm{R}+\bm{r}) \alpha, \bm{r}^\prime \gamma} \exp[{i\bm\theta\cdot(\bm{R}+\bm{r}-\bm{r}^\prime)}],
\end{equation} 
for a general matrix $A$ in real space. We find that $\Sigma_{i\omega}^{\bm\theta} = \Sigma_{i\omega}$ and   
${G}_{i\omega}^{\bm\theta} = 1/({i\omega \bm{1}-H_0^{\bm\theta}-\Sigma_{i\omega}}) $, with $\bm{1}$ being the identity matrix.
Then, we the GIMF can be rewritten as,
\begin{equation}
{\chi}=\frac{ \epsilon^{\tilde{\mu}\tilde{\nu}}}{8 \pi^2}  \int d\omega d{\bm \theta} {\rm Tr}[(\partial_{\tilde\mu} H_0^{\bm \theta}) {G}_{i\omega}^{\bm \theta} (\partial_{\tilde\nu} H_0^{\bm \theta}){G}_{i\omega}^{\bm \theta} A_{i\omega} {G}_{i\omega}^{\bm \theta}], \label{ChernNum31}
\end{equation}
where $\tilde\mu$ and $\tilde\nu$ run through $\theta_x$ and $\theta_y$ now, and $A_{i\omega} = i \bm{1} - \partial_\omega \Sigma_{i\omega}$. A detailed derivation can be found in the supplementary \cite{jzhengxx}.  Eq.\,\eqref{ChernNum31} is exactly the first Chern number of a periodic system with the Bloch momentum $\bm \theta$, in which the sites in the supercell are treated as internal degree of freedom, i.e., a pseudospin. The method is consistent with the proposal of a topological index in a disordered system via introducing twisted phases for the Green's function \cite{Wang2010prl}.

 Now we can prove that the localized states do not contribute to the Hall conductance. As shown in Refs. \cite{Shindou2006} and \cite{Zheng2017}, Eq.\,\eqref{ChernNum31} can be expressed by the summation of the Berry curvature of all occupied quasi-particle states \cite{jzhengxx}. Here, the quasi-particle state is  the eigenstate of $G^{\bm \theta}_\omega$ at its pole for real frequencies. By varying the twisted phase, we obtain quasi-particle bands in $\bm \theta$ space. Let us suppose that the size of the supercell is significantly larger than the localization length. Then for a localized state, the twisted boundary condition does not change the pole of the Green's function and the corresponding eigenstate apart from a unitary transformation, since the eigenstate vanishes at the boundary if we properly choose the position of  the edge of the supercell. Thus a localized state forms a flat band upon varying the twisted phase and the corresponding Berry curvature is always trivial \cite{jzhengxx}. This implies that localized states do not contribute to the conductance. 

The self-energy in the Green's function can be obtained by using real-space DMFT \cite{Hofstetter2008} for each disorder configuration in a FSS.  Within this approach, the system is mapped to a set of coupled single impurity problems, where the other sites, acting as a  bath, are integrated out. The effective action for each single impurity is
$S_{\bm{i}} = - \int d\tau   d\tau^{\prime} {c}^\dagger (\tau) \mathcal{G}_{0,\bm{i}}^{-1}(\tau-\tau^{\prime}){c}(\tau^{\prime})+{U}\int d\tau n_{\bm{i} \downarrow} n_{\bm{i}\uparrow}$,
where $\mathcal{G}_{0,\bm{i}}$ is the Weiss function at the site $\bm{i}$ and the spin indices are hidden \cite{Georges1996}. The full Green's function is given by the Dyson equation, $
\mathcal{G}_{\bm{i}}^{-1}(i\omega_{n})=\mathcal{G}_{0,\bm{i}}^{-1}(i\omega_{n})-{\Sigma}_{\bm{i}\bm{i}}(i\omega_{n})$, where  each term depends on the site $\bm{i}$ in the supercell. We use iterative perturbation theory to obtain the self-energy for each single impurity problem  \cite{Kajueter1996prl, Potthoff1997prb}. The full lattice Green's function is given by ${G}({i\omega_n}) = 1/(i\omega_n \bm{1} - H_0 - \Sigma)$.  A self-consistent solution is found by closing the loop with $ \mathcal{G}_{\bm{i}}  =G_{\bm{i}\bm{i}}$. 

In addition, for the case of weak disorder and interaction, we also develop the EMT 
\cite{jzhengxx}. It is a perturbative mean-field method, in which the  effective Hamiltonian $ H_{\rm{eff}} \equiv h_0+\overline{\Sigma}$ obtains a translationally invariant form. The self-energy is the statistically averaged result over different disorder samples. Specifically, we focus on a system with an on-site potential  $  \hat{H}_{\rm{dis}}= \sum_{\bm{i}\alpha} V_{\bm{i}}\hat n_{\bm{i}\alpha}$, where $V_{\bm i}$ is random and position-dependent but uniformly distributed in $[-W, W]$. The self-energy is defined by 
\begin{equation}
 \overline{G} = 1/{(\omega \bm{1} - h_0 -\overline{\Sigma})}
\end{equation}
at $\omega = 0$, where $ \overline{G}$ is the disorder-averaged Green's function.  We can prove that, the self-energy, up to the order of $W^2$ and $U$,  is given by
\begin{equation}
\overline\Sigma_{\bm{i}\alpha,\bm{i}\gamma} =  \overline\Sigma^{\rm{dis}}_{\bm{i}\alpha,\bm{i}\gamma} + \overline\Sigma^{\rm{U}}_{\bm{i}\alpha,\bm{i}\gamma},
\end{equation}
where $  \overline\Sigma^{\rm{dis}}_{\bm{i}\alpha,\bm{i}\gamma} = {W^2}  \overline{G}_{\bm{i}\alpha,\bm{i}\gamma}/ {3}$ 
is contributed by the disorder \cite{Groth2009prl} and $ \overline\Sigma^{U}_{\bm{i}\alpha,\bm{i}\gamma} $ is the statistical average of the Hartree -Fock self-energy $U\langle \hat{c}_{\bm{i}\alpha}\hat{c}_{\bm{i}\bar{\alpha}}^{\dagger}\rangle \delta_{\bar{\alpha}\gamma}+U\langle \hat{c}_{\bm{i}\bar{\alpha}}^{\dagger}\hat{c}_{\bm{i}\bar{\alpha}}\rangle\delta_{\alpha\gamma}$ with $\bar\uparrow = \downarrow$ and  $\bar\downarrow =\uparrow$. The statistical expectation values of the local operators $\hat{c}_{\bm{i}\alpha}\hat{c}_{\bm{i}\bar{\alpha}}^{\dagger}$ and $\hat{c}_{\bm{i}\bar{\alpha}}^{\dagger}\hat{c}_{\bm{i}\bar{\alpha}}$ can be calculated through the effective Hamiltonian $ H_{\rm{eff}}$.  Thus the self-energy can be solved self-consistently. This EMT cannot predict  Anderson localization. It describes the contribution of the extended states.  Later, we will see that the self-energy $\overline{\Sigma}$ can effectively describe  the band inversion  due to   interaction and disorder.

Now, we consider the HHHM on the square lattice  \cite{Hatsugai1990prb}.  The corresponding one-dimensional model after the dimensional reduction is the Rice-Mele model, the Thouless pumping of which has been observed in experiments for both bosonic and fermionic systems  \cite{Nakajima2016,Lohse2016}.  The translationally invariant part  of the Hamiltonian is 
\begin{eqnarray}
\hat{h}_0 &= & - \sum_{\bm{i},\alpha}  {\big [}t_x  \hat{c}^\dagger_{\bm{i}+\hat x,\alpha} \hat{c}_{\bm{i}\alpha} +  t_y e^{i 2\pi \xi i_x}\hat{c}^\dagger_{\bm{i}+\hat y,\alpha} \hat{c}_{\bm{i} \alpha}  + {\rm{ H.c.}} {\big ]} \notag \\
 &&-\sum_{\bm{i},\alpha}  t_{z} {\big [} e^{i 2\pi \xi (\bm{i}_x+\frac{1}{2})}  
(\hat{c}^\dagger_{\bm{i}+\hat x+\hat y,\alpha} \hat{c}_{\bm{i}\alpha} + \hat{c}^\dagger_{\bm{i}+\hat y,\alpha} \hat{c}_{\bm{i}+\hat x,\alpha}) \notag \\ &&
+~ {\rm{ H.c.}} {\big]} + \sum_{\bm{i},\alpha} \left[ (-1)^{{i}_x} \Lambda  - \mu_0 \right] \hat n_{\bm{i}\alpha} . \label{h0}
\end{eqnarray}
Here, $t_x$ and $t_y$ are the nearest-neighbor hopping along the $x$ and $y$ direction respectively,  $t_z$ is the next-nearest-neighbor hopping, $\mu_0$ is the chemical potential determined by the filling, $\hat{x} =(1,0)$ and $\hat{y} = (0,1)$ are the unit vectors and $\bm{i} = (i_x, i_y)$.  We focus on the  case with $\xi =1/2$, in which there is a $\pi$-flux in each unit square.  The total particle number  is fixed to be $N$, so that the average filling is $1/2$. 

 Without disorder and interaction,  since  the spin is conserved and due to the $SU(2)$ symmetry in spin space, the Hamiltonian becomes $h_{0} = \bm{v}(\bm{k})\cdot \bm{\sigma} -\mu_0$  in momentum space. Here,  $\bm{v}(\bm{k})= -(2 t_x \cos k_x,  4 t_z \sin k_y  \sin k_x, \Lambda - 2t_y \cos k_y )$ is  the Bloch vector and $\bm \sigma$ are the Pauli matrices in the pseudospin corresponding to  odd (up) and even (down) position along the $x$-axis. The identity matrices in spin and pseudospin spaces are omitted.   For $|\Lambda|< 2t_y$,   the Bloch vector covers the origin,  and thus, the system is a topological insulator. Increasing the staggered potential, the system closes the band gap at $|\Lambda| = 2t_y$, it becomes  a trivial insulator thereafter. 

For the weakly disordered and interacting case, we give a discussion based on the EMT. In the EMT, the effective Hamiltonian is  $H_{\rm{eff}} = h_0 + \overline\Sigma$, where $\overline\Sigma = \overline\Sigma_0 + \overline{ \Sigma}_z {\sigma_z}$. For the half filling case, with a small  but finite temperature, we have $\mu_0 = \overline\Sigma_0$ due to the particle-hole symmetry. Thus, the effective Hamiltonian becomes $H_{\rm{eff}}= \bm{d}\bm{\cdot} \bm{\sigma}$, where ${d}_x = \bm{v}_x$, $d_y=\bm{v}_y$ and $d_z = \bm{v}_z+ \overline{\Sigma}_z$.  From the EMT, we get  $\overline\Sigma^{\rm{dis}} = -\frac{W^2}{12 \pi^2} \int d^2 \bm{k} ~\frac{{d}_z {\sigma_z}}{|\bm{d}|^2}  $ and $\overline\Sigma^{\rm U} = \frac{U}{2} (n_{\rm o}+n_{\rm e}) +  \frac{U}{2}   (n_{\rm o}-n_{\rm e}) \sigma_z$, where $n_{\rm o}$ and $n_{\rm e}$ is the particle number for each spin at the odd and the even position along the $x$-axis. The integrate region is $k_x, k_y \in [0, 2\pi]$, which is the doubling of the Brillouin zone. The Fock term does not appear in $\overline\Sigma^{\rm U}$ since the spin is conserved. At half filling, for each spin, we have $n_{\rm o}+n_{\rm e}=1$.  Note that $n_{\rm o}-n_{\rm e} = \langle \sigma_z \rangle$ is the pseudospin polarization. Thus, we obtain the chemical potential  $\mu_0 = U/2 $ and the  self-consistent self-energy
\begin{equation}
\overline\Sigma = \frac{U}{2} -\left\{\frac{W^2}{12 \pi^2} \int d^2{\bm k} \frac{{d}_z }{|\bm{d}|^2} + \frac{U}{8 \pi^2} \int d^2{\bm k} \frac{{d}_z } {|\bm{d}|} \right\} {\sigma}_z. \label{self-energy}
\end{equation}

 As an approximation, one can use $\bm{v}$ to replace $\bm{d}$ in the integral above. The effective staggered potential in $H_{\rm{eff}}$ then becomes $\Lambda_{\rm{eff}} =\Lambda - \delta\Lambda$, where $ \delta\Lambda= - \frac{W^2}{12 \pi^2} \int d^2{\bm k} \frac{\bm{v}_z }{|\bm{v}|^2} - \frac{U}{8 \pi^2} \int d^2{\bm k} \frac{\bm{v}_z } {|\bm{v}|}$. Independently of the sign of $\Lambda$, the weak disorder and interaction effectively smoothen the  staggered potential. Intuitively, one would expect that the disorder may randomly intensify or weaken the staggering. However, states are preferably localized around those positions where the staggered potential has been increased. Thus for the extended states, which contribute to the conductance, the effective staggered potential becomes smooth. In addition, the repulsive interaction prefers to induce a uniform distribution of particles in space. This explains why the interaction assists the disorder in decreasing the effective  staggered potential. 

\begin{figure}
 \includegraphics[width=0.95\columnwidth]{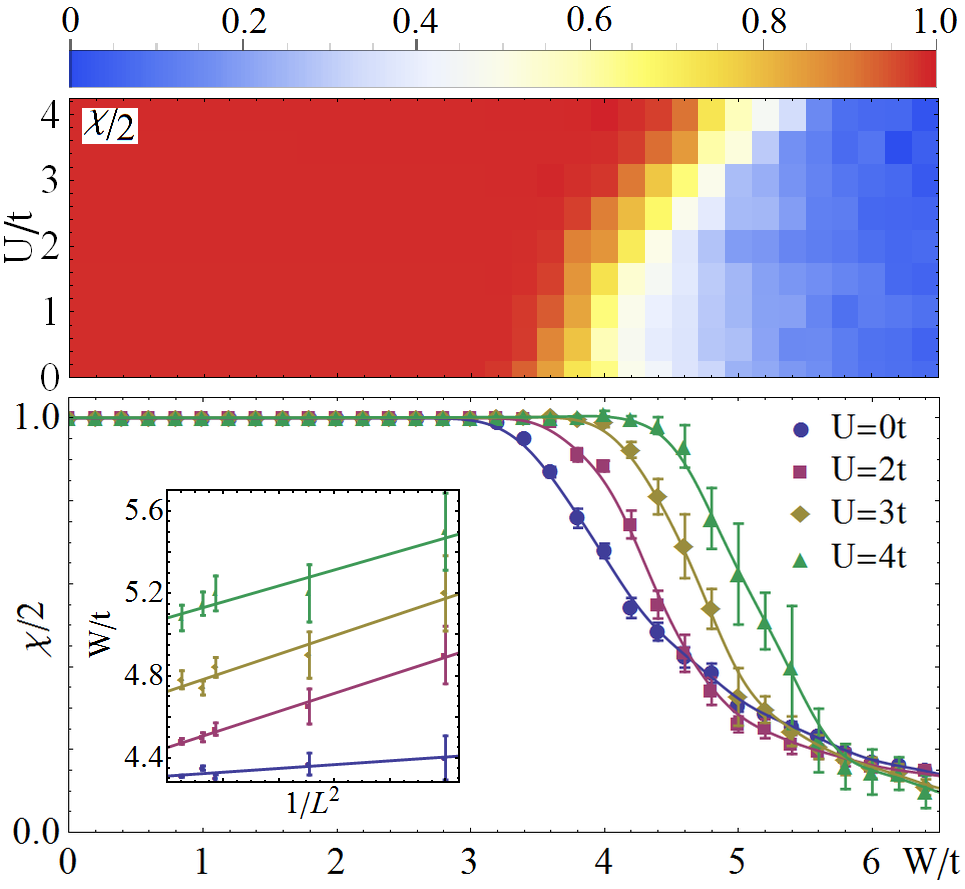} 
\caption{The density plot (upper) and errorlistplot (lower) of $\chi/2$ as a function of disorder strength and interaction strength, for $\Lambda =0$. The inset shows  the disorder strength where $\chi/2 = 0.5$, as a function of $1/L^2$ and for different interaction strengths, where $L^2$ is the size of the supercell with $n_x = n_y = L$ (from left to right: $ L = 44, 28, 24, 14, 10$). }
\label{fig1} 
\end{figure}

\begin{figure}
 \includegraphics[width=0.95\columnwidth]{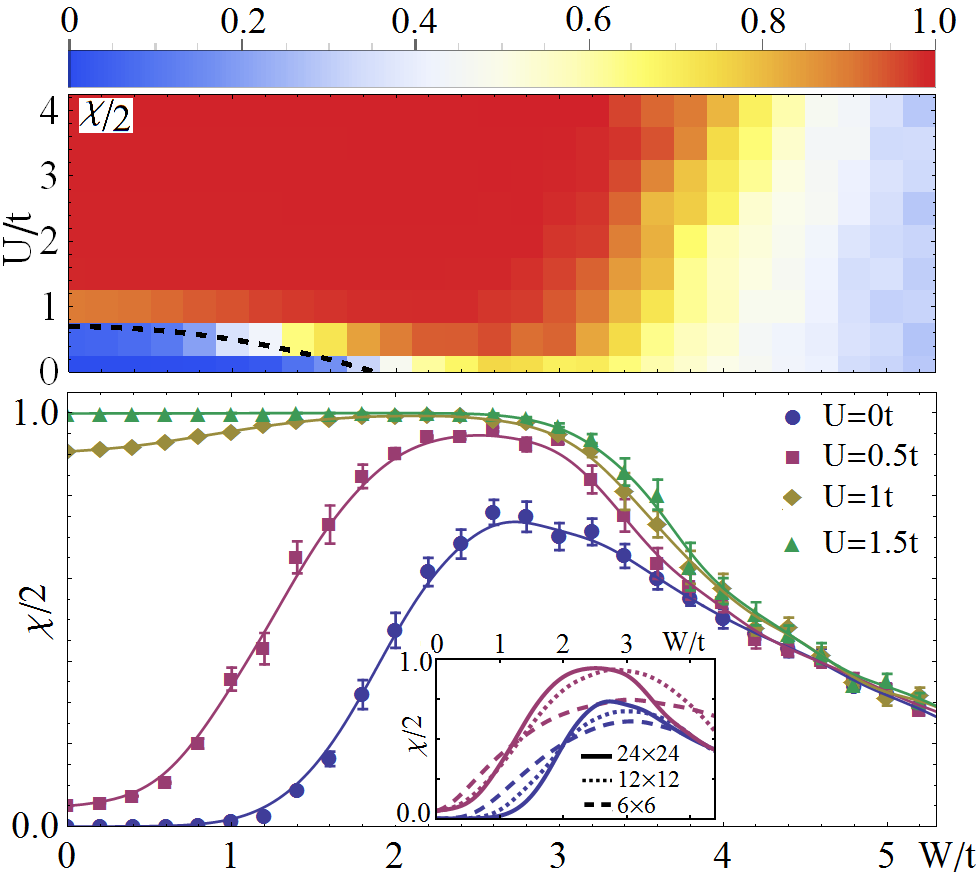} 
\caption{The density plot (upper) and errorlistplot (lower) of $\chi/2$ as a function of disorder strength and interaction strength, for $\Lambda = 2.2t$. The dashed black line shown in the upper figure is given by the EMT,  see Eq.\,\eqref{self-energy}. The inset in the lower figure shows how the value of $\chi/2$ depends on the size of supercell, where the blue (red) lines are for $U = 0.0 t ~(0.5 t)$. }
\label{fig2} 
\end{figure}

For the general case of arbitrary interaction- and disorder- strengths, we use DMFT to describe interaction effects. In Fig.\,\ref{fig1} and Fig.\,\ref{fig2}, we show our numerical results for the case $n_x=n_y=24$. For simplicity, we choose parameters $t_x =t_y =t_z = t$ and  $ k_B T =t/40 $. The grid for the twisted phase $\bm \theta$ is $8\times 8$.
For each value of $U$ and $W$, we generate $5 - 100$ sample realizations of the disorder. The final Hall conductance is obtained by averaging the Hall conductance over these samples. Numerically, deviations of the obtained conductance from the quantized value, are mainly due to finite size effects and partially due to finite temperature effects.

In  Fig.\,\ref{fig1},  we show the DMFT results for the Hall conductance for a system without staggered potential. For $W=U =0$, the system is a topological insulator. In the EMT, the effective Hamiltonian is not changed by interaction and disorder, i.e., $\Lambda_{\rm{eff}} =\Lambda = 0$. This means that no band inversion occurs. However, disorder broadens the distribution of the spectrum in each of the  two bands. The marginal states of each band become localized. In the DMFT calculation, by increasing the disorder strength,  the Hall conductance, as a topological invariant, is well quantized before the gap is closed. Thereafter  the Hall conductance decreases when further increasing the disorder, until all of the extended states become localized. From  Fig.\,\ref{fig1}, we observe that the interaction effectively impedes the closing of gap and the subsequent Anderson localization. In the inset, we also plot the critical disorder strength (where $\chi/2 = 0.5$) as a function of $1/L^2$ and for different interaction strengths, where $n_x = n_y = L$ is the size of the supercell.  The scalling behavior shows an approximately linear dependence of the critical  disorder strength on  $1/L^2$.

  In  Fig.\,\ref{fig2}, we show the DMFT result of the Hall conductance for  $\Lambda = 2.2 t$. The system is a topologically trivial insulator for $W=U=0$. For weak disorder and weak interaction, both increasing the disorder strength and increasing the interaction strength will raise the Hall conductance. From EMT we know that the repulsive on-site interaction can assist the disorder to smoothen the large staggered potential more effectively, and induce the integer quantum Hall effect through band inversion, which occurs at $\Lambda_{\rm eff} = 2.0 t$. Comparing with the DMFT results, the EMT gives an accurate prediction for the topological phase transition line for weak disorder and interaction (see the dashed black line shown in the density plot of $\chi/2$ in Fig.\,\ref{fig2}). When disorder becomes stronger,  the Hall conductance decreases again. The interaction also impedes formation of Anderson localization through  many-body effects. In the inset, we show $\chi/2$ for different sizes of the supercell. For weak disorder, the intersection points of different curves approximately predict the location of the phase transition. From the inset, we observe that the phase transition is located around $W=2.0 t$ for $U=0$ and around $W=1.0 t$ for $U=0.5t$.

In both cases, the interaction broadens the regime of the topological phase in the disordered system.  The interaction effectively smoothens the staggered potential and has a delocalizing effect.  The prediction from EMT deviates from the DMFT results when disorder becomes large. Note that quenched disorder is generically present in electronic condensed matter, and can also be realized in cold atom systems \cite{Damski2003prl,Lye2005prl,Clement2005prl,Clement2006NJP,White2009prl,Kondov2015prl}.
The one-dimensional topological Anderson insulator has been obsevered in experiment \cite{Meier2018}. In addition, the Chern number has been measured in the Harper-Hofstadter model through the drift of the atomic cloud \cite{Aidelsburger2015NP}. For the HHHM, an additional next-nearest-neighbor hopping is needed. 

In summary, we have generalized the Ishikawa-Matsuyama formula for the topological index to systems without translational invariance and calculated  the Hall conductance for a lattice system in the presence of disorder, within the dynamical mean-field approximation. The combined effect of interaction and disorder is discussed.  We find that the integer quantum Hall effect is enhanced by the on-site interaction for disordered systems.

\begin{acknowledgments}
{\it Acknowledgments}: J.-H. Zheng would like to thank Lijia Jiang and Bernhard Irsigler for careful  reading of the draft. This work was supported by the Deutsche Forschungsgemeinschaft via DFG FOR 2414, DFG HO 2407/9-1 and the high-performance computing center LOEWE-CSC.
\end{acknowledgments}

\newpage
\begin{widetext}
\noindent{\LARGE \textbf{Supplementary for ``Interaction-enhanced \\
 integer quantum  Hall effect in  disordered systems"}}
~\\

\large
In this supplementary, we show the details for the Hall conductance in disordered systems and for the effective medium theory.

\section{Hall Conductance}
{\bf Current operators}:
In the Heisenberg picture, the continuity equation for the particle number is \cite{Bernevig2013book}
\begin{equation}
\dot{\hat\rho}+\nabla\cdot \hat{ \bm J}=0.
\end{equation}
After Fourier transformation this becomes $\dot{\hat\rho}_{\bm{q}}-i\bm{q\cdot } \hat{\bm J}_{\bm{q}}=0$.
Using the local density operator $\hat\rho_{\bm i}= \sum_\alpha \hat c_{\bm i \alpha}^{\dagger}\hat c_{\bm i \alpha}$,  we obtain $\hat \rho_{\bm{q}}  =\frac{1}{\sqrt{N}}\sum_{\bm{ k}} \hat c_{(\bm{k+q}) \alpha}^{\dagger}\hat c_{\bm{k} \alpha}$.  By using the equation of motion for the density operator, 
the continuity equation yields 
\begin{equation}
-\bm{q\cdot} \hat{\bm J}_{\bm{q}}=i\dot{\hat \rho}_{\bm{q}}=\big[\hat\rho_{\bm{q}},\hat H\big].  \label{continue}
\end{equation}
Since the on-site interaction term is locally particle-number conserving, it is easy to check $\big[\hat\rho_{\bm{i}},\hat H_{\rm{int}}\big] = 0 $.   For the non-interacting part $\hat{H}_0 = \sum_{\bm{k}_1\alpha,\bm{k}_2\gamma}\hat{c}_{\bm{k}_1\alpha}^{\dagger}[H_0]_{\bm{k}_1\alpha,\bm{k}_2\gamma}\hat{c}_{\bm{k}_2\gamma}$, we have 
\begin{eqnarray}
\big[\hat\rho_{\bm{q}},\hat H_{0}\big]  & =&  \frac{1}{\sqrt{N}} \sum_{\bm{k}_1\alpha,\bm{k}_2\gamma} [H_0]_{\bm{k}_1\alpha,\bm{k}_2\gamma}
\left\{\hat{c}_{(\bm{k}_1 + \bm q )\alpha}^{\dagger}\hat{c}_{\bm{k}_2\gamma} -  \hat{c}_{{\bm{k}_1  }\alpha}^{\dagger}\hat{c}_{(\bm{k}_2 - \bm q)\gamma}\right\}  \notag \\
&= & \frac{1}{\sqrt{N}} \sum_{\bm{k}_1\alpha,\bm{k}_2\gamma}
\left\{  [H_0]_{(\bm{k}_1- {\bm q}/2 )\alpha,(\bm{k}_2 - {\bm q}/2 )\gamma} -  [H_0]_{(\bm{k}_1+{\bm q}/2 )\alpha, (\bm{k}_2+ {\bm q}/2 )\gamma}      \right\} \hat{c}_{(\bm{k}_1 + {\bm q}/2 )\alpha}^{\dagger}\hat{c}_{(\bm{k}_2 - {\bm q}/2 )\gamma}.
\end{eqnarray}%
Thus, for small $\bm{q}$,  from Eq.\eqref{continue} we obtain the current operator,
\begin{equation}
\hat{ \bm J}_{\bm{q}}=\frac{1}{\sqrt{N}}\sum_{\bm{k}_1\alpha,\bm{k}_2\gamma} \hat c_{(\bm{k}_1+{\bm{q}}/{2})\alpha}^{\dagger}[\nabla_{\bm{k}_1}H_{0}({\bm{k}_1 \alpha, \bm{k}_2 \gamma})+\nabla_{\bm{k}_2}H_{0}({\bm{k}_1 \alpha, \bm{k}_2}\gamma)] {\hat c}_{(\bm{k}_2-{\bm{q}}/{2})\gamma}+\mathcal{O}(\bm{q}).
\end{equation}

{\bf Hall conductance}: The Kubo formula for the dc Hall conductance is \cite{Mahan2000book}
\begin{equation}
\sigma_{H}=\lim_{\omega\rightarrow0}\frac{1}{\omega}{\rm{Im}}\, \Pi_{yx}(i\omega_{n}\rightarrow\omega+i 0^+),
\end{equation}
where the current-current correlation function is
\begin{equation}  \Pi_{yx}(i\omega_{n})=\lim_{\bm{q\rightarrow 0}}\int_{0}^{\beta}d\tau e^{i\omega_{n}\tau}\langle T_{\tau}{\hat{J}_{\bm{q}}^{y}}(\tau){{\hat{J}_{-\bm{q}}^{x}}(0)}\rangle. \label{kubo}
\end{equation} 
Using L\textsc{\char13}H\^opital's rule and the Cauchy-Riemann equations, the dc Hall conductivity becomes
\begin{equation}  
\sigma_{H}= \partial_{\omega_x} {\rm{Im}}\, \Pi_{yx}(z)|_{z = i 0^+}  =  -\partial_{\omega_y}{\rm{Re}}\,\Pi_{yx}(z)|_{z = i 0^+} =  -\partial_{\omega}{\rm{Re}}\,\Pi_{yx}(i \omega)|_{\omega = 0^+} .
\end{equation}
where $z =\omega_x +i \omega_y$.

As shown in the main text, within DMFT, the correlation function \eqref{kubo} becomes
\begin{equation}
\Pi_{yx}(i\omega_{n})= \frac{1}{N \beta}\sum_{n^{\prime}}{\rm Tr} [\bar\partial_{k_y}  {G}^{-1}(i\omega_{n^{\prime}}) {G}(i\omega_{n^{\prime}}+i\omega_{n})\bar\partial_{k_x}  {G}^{-1}(i\omega_{n^{\prime}}) {G}(i\omega_{n^{\prime}})].
\end{equation}
In the zero temperature limit, we can use the replacement $\frac{1}{\beta}\sum_{n^{\prime}} \rightarrow \int \frac{d\omega}{2\pi}$, and thus the summation becomes an integral. In addition, using the fact that $\partial_\omega G = - G \partial_\omega {G}^{-1} G$, we obtain
\begin{equation}
F_{\mu\nu} \equiv  -\partial_{\omega}\Pi_{\mu\nu}(i\omega)|_{\omega = 0^+} = \frac{1}{2\pi N} \int d\omega {\rm Tr}[{G}_{i\omega}\bar\partial_{k_\mu}{G}_{i\omega}^{-1}{G}_{i\omega}\partial_{\omega}{G}_{i\omega}^{-1}{G}_{i\omega}\bar\partial_{k_\nu}{G}_{i\omega}^{-1}], \label{ChernNum}
\end{equation}
for $\mu, \nu = x, y$.
Note that the Hall conductance is defined as $\sigma_h = {\rm{Re}}\,F_{yx} = (F_{yx} +F_{yx}^*)/2$.  From the Lehmann representation, we can find  that the Green's function satisfies the following relation
\begin{equation}
(G^{-1}_{i\omega})^* =  (G^{-1}_{-i\omega})^T,
\end{equation}
where $T$ means transpose. It implies $(\bar\partial_{k_\mu} G^{-1}_{i\omega})^* =  \bar\partial_{k_\mu} (G^{-1}_{-i\omega})^T = \bar\partial_{k_\mu} (G^{-1}_{i \tilde\omega})^T |_{\tilde\omega = -\omega}$ and $(\partial_\omega G^{-1}_{i\omega})^* =  \partial_\omega (G^{-1}_{-i\omega})^T = - \partial_{\tilde\omega} (G^{-1}_{i \tilde\omega})^T |_{\tilde\omega = -\omega}  $.  Using the fact that a matrix and its transpose have the same trace, we can obtain $\sigma_h = (F_{yx} - F_{xy})/2$.
Since the trace is invariant under cyclic permutations,  we finally obtain
\begin{equation}
{\chi}\equiv \frac{\epsilon^{\tilde{\mu}\tilde{\nu}\tilde{\rho}}}{6N}\int d\omega {\rm Tr}[{G}_{i\omega}(\bar\partial_{\tilde{\mu}}{G}_{i\omega}^{-1}){G}_{i\omega}(\bar\partial_{\tilde{\nu}}{G}_{i\omega}^{-1}){G}_{i\omega}(\bar\partial_{\tilde{\rho}}{G}_{i\omega}^{-1})], \label{ChernNum}
\end{equation}
where the indices $\tilde{\mu}$, $\tilde{\nu}$, and $\tilde{\rho}$ run through $k_x$, $k_y$, and $\omega$, and   $ [\bar\partial_{\omega}{G}_{i\omega}]_{\bm{k}_{1}\alpha,\bm{k}_{2}\gamma}\equiv\partial_{\omega}{G_{i\omega}}(\bm{k}_{1}\alpha,\bm{k}_{2}\gamma) $.

{\bf Hall conductance as a topological index}: For any two matrices ${A}$ and ${B}$ in $\bm{k}$-space, according to our convention, we have 
\begin{equation}
\left[\bar\partial_{k_\mu}({AB})\right]_{\bm{k}_{1},\bm{k}_{2}} = (\partial_{k_{1, \mu}}+\partial_{k_{2, \mu}})\sum_{\bm{k}_3}{A}(\bm{k}_{1},\bm{k}_3){B}(\bm{k}_3,\bm{k}_{2}).
\end{equation} 
In addition, we also have $ \sum_{\bm{k}_3}\partial_{k_{3,\mu}}\{{A}(\bm{k}_{1},\bm{k}_3){B}(\bm{k}_3,\bm{k}_{2})\} =0 $, which is obvious if we use the fact $\sum_{\bm{k}}\propto \int d\bm{k}$. Here,  the spin indices $\alpha$ and $\gamma$ are omitted for notation simplicity. The above two equations show that the derivative operator $\bar\partial_{k_\mu}$ satisfies the Leibniz product rule,
\begin{equation}\bar\partial_{k_\mu}({AB})=(\bar\partial_{k_\mu}{A}){B}+{A}\bar\partial_{k_\mu}{B}.
\end{equation}
Besides, using the fact that ${G}_{i\omega}{G}_{i\omega}^{-1}=\bm{1}$ (for well defined $G$ and $G^{-1}$),
we have  
\begin{equation}
\delta({G}_{i\omega}\bar\partial_{k_\mu}{G}_{i\omega}^{-1})=-{G}_{i\omega}\left(\bar\partial_{k_\mu}{G}_{i\omega}^{-1}\right)\delta{G}_{i\omega}{G}_{i\omega}^{-1}-\left(\bar\partial_{k_\mu}\delta{G}_{i\omega}\right){G}_{i\omega}^{-1}.
\end{equation}
 With these preparations, similar in the Ref.  \cite{Qi2008prb}, it is not difficult to check that 
\begin{equation}
{\delta \chi [{G}_{i\omega}]} \varpropto  {\epsilon ^{\tilde\mu \tilde\nu\tilde \rho }}\int d\omega {\rm Tr} [\bar \partial _{\tilde\mu
} ( {G}_{i\omega}^{-1}\delta {G}_{i\omega}\bar\partial _{\tilde\nu }{G}_{i\omega}^{-1}%
{G}_{i\omega}\bar\partial _{\tilde\rho }{G}_{i\omega}^{-1}{G}_{i\omega})]  =0.
\end{equation}
Thus, $\chi$ is invariant under infinitesimal deformations of the Green's function, and therefore is a topological index.

\section{Hall conductance in the extended infinite system}
The position $\bm{i}$ in the EIS can be  expressed as $\bm{i}=\bm{R}+\bm{r}$, where $\bm{R}$ refers to the position of the corresponding supercell and $\bm{r}$ is the relative position in the supercell.  For a general matrix $A$ in real space, the translational symmetry implies the relation,
$A_{ (\bm{R}+\bm{r})\alpha,(\bm{R}^\prime+\bm{r}^\prime) \gamma}=A_{(\bm{R}-\bm{R}^\prime+\bm{r})\alpha, \bm{r}^\prime\gamma }$. 
Using the Fourier transformation, in momentum space, we have 
\begin{equation}
A_{\bm{k}\alpha,\bm{k^\prime}\gamma} \equiv \frac{1}{N}  \sum_{\bm{R}\bm{R}^\prime \bm{r}\bm{r}^\prime}  A_{(\bm{R}+\bm{r})\alpha, (\bm{R}^\prime+\bm{r}^\prime) \gamma} e^{i \bm{k}\cdot (\bm{R}+\bm{r}) -i \bm{k}^\prime\cdot (\bm{R}^\prime+\bm{r}^\prime)},
\end{equation}
for $k_\mu, k_\mu^{\prime} \in [0, 2\pi)$ with $\mu =x, y$, where the lattice constant is set to be $a =1$. The EIS retains the translational symmetry at a large scale, and the Bloch wavevector is denoted as $\theta_\mu \in [0, 2\pi/n_\mu)$, where $n_x\times n_y$ is the size of the supercell. Splitting the momenta as $k_\mu= \tilde k_\mu + \theta_\mu$ and $k_\mu^{\prime}= \tilde k_\mu^{\prime} + \theta_\mu^{\prime}$, with $\tilde k_\mu = 2 \pi l /n_\mu $ and $ \tilde k_\mu^{\prime} = 2 \pi l^{\prime}  /n_\mu $ $(l,l^{\prime}  =0, 1, \cdots, n_\mu-1)$, and using the fact that $e^{i {\bm{\tilde{k}}\bm{\cdot R}}}=e^{i {\bm{\tilde{k}^{\prime}}\bm{\cdot R}^{\prime}}}=1$, we obtain
\begin{equation}
A_{\bm{k}\alpha,\bm{k^\prime}\gamma} = \frac{1}{N}  \sum_{\bm{R}\bm{R}^\prime \bm{r}\bm{r}^\prime}  A_{(\bm{R} -\bm{R}^\prime +\bm{r})\alpha, \bm{r}^\prime \gamma} e^{i \bm{\theta}\cdot (\bm{R}+\bm{r}) -i \bm{\theta}^\prime\cdot (\bm{R}^\prime+\bm{r}^\prime)}e^{i \bm{\tilde k}\cdot \bm{r} -i \bm{\tilde k}^\prime\cdot \bm{r}^\prime }.
\end{equation}
Introducing new variables $\bm{R}_s \equiv \bm{R}^\prime$ and $\bm{R}_a \equiv \bm{R} - \bm{R}^\prime$, and using the fact that $\sum_{\bm{R}_s} e^{i (\bm{\theta}-\bm{\theta}^\prime)\bm{\cdot}\bm{R}_s} = \frac{N}{n_x n_y} \delta_{\bm \theta \bm\theta^\prime} $, we obtain
\begin{equation}
A_{(\tilde{\bm{k}}+\bm{\theta})\alpha,(\tilde{\bm{k}}^\prime+\bm{\theta}^\prime)\gamma} = \frac{1}{n_x n_y}  \sum_{\bm{R}_a\bm{r}\bm{r}^\prime}  A_{(\bm{R}_a +\bm{r})\alpha, \bm{r}^\prime \gamma} e^{i \bm{\theta}\cdot (\bm{R}_a+\bm{r}-\bm{r}^\prime)}e^{i \bm{\tilde k}\cdot \bm{r} -i \bm{\tilde k}^\prime\cdot \bm{r}^\prime }  \delta_{\bm \theta \bm\theta^\prime}.
\end{equation}
So that we have  
\begin{equation}
A_{ \bm{k}\alpha,\bm{k^\prime}\gamma} = A_{( \bm{\tilde k}+\bm{\theta})\alpha,(\bm{\tilde k^\prime}+\bm{\theta})\gamma} \delta_{\bm \theta \bm\theta^\prime}, \label{11}
\end{equation} 
which is exactly due to the translational symmetry. It also implies that  
\begin{equation}\bar\partial_{\mu} A_{ \bm{k}\alpha,\bm{k^\prime}\gamma} = [\partial_{\theta_\mu}A_{ (\bm{\tilde k}+\bm{\theta})\alpha,(\bm{\tilde k^\prime}+\bm{\theta})\gamma}] \delta_{\bm \theta \bm\theta^\prime}. \label{12}
\end{equation}
 Let us define 
\begin{equation} A^{\bm\theta}_{ \bm{r}\alpha, \bm{r}^\prime\gamma}\equiv \sum_{\bm{R}} A_{ (\bm{R}+\bm{r}) \alpha, \bm{r}^\prime\gamma} e^{i\bm\theta\cdot(\bm{R}+\bm{r}-\bm{r}^\prime)}, \label{twist}
\end{equation} 
and then 
\begin{equation}
A_{ (\bm{\tilde k}+\bm{\theta})\alpha,(\bm{\tilde k^\prime}+\bm{\theta})\gamma} = \frac{1}{n_x n_y}  \sum_{\bm{r} \bm{r}^\prime} A^{\bm\theta}_{ \bm{r}\alpha, \bm{r}^\prime\gamma} e^{i \bm{\tilde k}\cdot \bm{r} -i \bm{\tilde k}^\prime\cdot \bm{r}^\prime } \label{13}
\end{equation}
is exactly the Fourier transformation of $A^{\bm\theta}_{ \bm{r}\alpha, \bm{r}^\prime\gamma}$ in a supercell.

In the following, we would like to show that $A^{\bm\theta}_{ \bm{r}\alpha, \bm{r}^\prime\gamma}$ is corresponding to the matrix with twist phases. Denote the right eigenvectors  of $A^{\bm\theta}_{\bm{r}\alpha, \bm{r}^\prime\gamma}$ as $\psi^{\bm \theta}$ with an eigenvalue $E$, then we have 
\begin{equation}
\sum_{\bm{r}^\prime\gamma}A^{\bm\theta}_{ \bm{r}\alpha, \bm{r}^\prime\gamma} \psi^{\bm \theta}_{\bm{r}^\prime\gamma} = \sum_{\bm{R}^\prime\bm{r}^\prime \gamma} A_{ (\bm{R}+\bm{r}) \alpha, (\bm{R}^\prime+\bm{r}^\prime)\gamma} e^{i\bm\theta\cdot(\bm{R}-\bm{R}^\prime+\bm{r}-\bm{r}^\prime)}  \psi^{\bm \theta}_{\bm{r}^\prime\gamma} = E  \psi^{\bm \theta}_{\bm{r}\alpha}.
\end{equation}
And thus 
\begin{equation}
 \sum_{\bm{R}^\prime\bm{r}^\prime \gamma} A_{ (\bm{R}+\bm{r}) \alpha, (\bm{R}^\prime+\bm{r}^\prime)\gamma} e^{-i\bm\theta\cdot(\bm{R}^\prime+\bm{r}^\prime)}  \psi^{\bm \theta}_{\bm{r}^\prime\gamma} = E   e^{-i\bm\theta\cdot(\bm{R}+\bm{r})}  \psi^{\bm \theta}_{\bm{r}\alpha}.
\end{equation}
So the corresponding eigenstates of $A_{ (\bm{R}+\bm{r})\alpha,(\bm{R}^\prime+\bm{r}^\prime) \gamma}$ becomes $\phi_{(\bm{R}+\bm{r})\alpha} \equiv e^{-i{\bm \theta}\cdot (\bm{R}+\bm{r})} \psi^{\bm\theta}_{\bm{r}\alpha}$, which shows the relation at the boundary of supercell,  $\phi_{(\bm{r} + n_\mu \hat{\mu}) \alpha} = e^{-i{ \theta_\mu} n_\mu} \phi_{\bm{r}\alpha}$. The phase  $\theta_\mu n_\mu$ is an effective twisted phase boundary condition in the supercell. 

 Applying this definition \eqref{twist} for the Hamiltonian $H_0$,  self-energy and the Green's function,  we find that $\Sigma_{i\omega}^{\bm\theta} = \Sigma_{i\omega}$ and   
${G}_{i\omega}^{\bm\theta} = 1/({i\omega \bm{1}-H_0^{\bm\theta}-\Sigma_{i\omega}}) $. After combining these results Eqs.\,\eqref{11}-\eqref{13},  the conductance  Eq.\,(\ref{ChernNum}) becomes
\begin{equation}
\sigma_{H}= \frac{\epsilon^{\tilde\mu\tilde\nu\tilde\rho}}{12\pi N}   \int d\omega  \sum_{\bm \theta} {\rm Tr}\{{G}_{i\omega}^{\bm \theta}\partial_{\tilde\mu}[{G}_{i\omega}^{\bm \theta}]^{-1}{G}_{i\omega}^{\bm \theta}\partial_{\tilde\nu}[{G}_{i\omega}^{\bm \theta}]^{-1}{G}_{i\omega}^{\bm \theta}\partial_{\tilde\rho}[{G}_{i\omega}^{\bm \theta}]^{-1}\}, \label{xxx}
\end{equation}
where  $\tilde\mu$, $\tilde\nu$, and $\tilde\rho$ run through $\theta_x$, $\theta_y$, and $\omega$ now and the trace is only for the supercell lattice and spin index. Using the fact that $\partial_{\tilde\mu}[{G}_{i\omega}^{\bm \theta}]^{-1} = - \partial_{\tilde\mu} H_0^{\bm \theta}$ for $\tilde\mu = \theta_x,\theta_y$ and $\partial_{\omega}[{G}_{i\omega}^{\bm \theta}]^{-1} =  i \bm{1} - \partial_\omega \Sigma_{i\omega}$, and using the replacement $\frac{1}{N}\sum_{\bm \theta} = \frac{1}{4 \pi^2} \int d\bm\theta$, we obtain $\sigma_H = {\chi}/{2\pi}$, where
\begin{equation}
{\chi}=\frac{ \varepsilon^{\tilde\mu\tilde\nu}}{8 \pi^2}  \int d\omega d{\bm \theta} {\rm Tr}[{G}_{i\omega}^{\bm \theta} (\partial_{\tilde\mu} H_0^{\bm \theta}) {G}_{i\omega}^{\bm \theta} (\partial_{\tilde\nu} H_0^{\bm \theta}){G}_{i\omega}^{\bm \theta} A_{i\omega} ]. \label{ChernNum3}
\end{equation}
Here, $\tilde\mu$ and $\tilde\nu$ run through $\theta_x$ and $\theta_y$ now, and $A_{i\omega} = \partial_\omega [{G}_{i\omega}^{\bm\theta} ]^{-1} = i \bm{1} - \partial_\omega \Sigma_{i\omega}$.  The expression \eqref{ChernNum3} is exactly the Chern number of a periodic system with the Bloch momentum $\bm \theta$, and the site index in the supercell as an internal degree of freedom like a pseudospin. Our method is consistent with the proposal for the Chern number in a disordered system via introducing twisted phases for the Green's function \cite{Wang2010prl}.

{\bf Berry curvature  of quasi-particle state and localized states}: The GIMF can be expressed as the summation of the Berry curvature of all occupied quasi-particle states, following the contour-integration method developed in Refs. \cite{Shindou2006,Zheng2017}.  The GIMF becomes
\begin{equation}
\chi=\sum_{\rm{occu.}}\frac{ \varepsilon^{\tilde\mu\tilde\nu}}{2\pi i}\int d^{2}{\bm \theta}\langle\partial_{\tilde\mu}\psi^{\bm \theta}|\partial_{\tilde\nu}\psi^{\bm \theta}\rangle,\label{eq:berry_ful}
\end{equation}
where $\psi^{\bm \theta}$ is the eigenvalue of ${G}_{\omega}^{\bm\theta}$ at its poles (for real frequencies). It is also the eigenvalue of $\omega \bm{1}-H_0^{\bm\theta}-\Sigma_{\omega} $ at the zeros. Considering a localized state $\psi$ for $\bm\theta=0$, we can shift the position of the supercell so that the localized state is almost in the center of the supercell and thus $\psi$ vanishes away from the center. Note the difference between $\omega \bm{1}-H_0^{\bm\theta}-\Sigma_{\omega} $ and $\omega \bm{1}-H_0^{\bm\theta = 0}-\Sigma_{\omega}$, is only the unitary transformation $e^{i \bm{\theta\cdot r}}$ besides the elements near the edge of supercell. However, for the localized state, the difference at on the edge of supercell has no effect,  and thus 
$\psi^{\bm \theta} = e^{i \bm{\theta\cdot r}}\psi^{\bm \theta =0}$. This kind of wave function contributes nothing to Eq. \eqref{eq:berry_ful}.

\section{Effective medium theory}
In the following, we derive the self-energy of a system with weak disorder and interaction, within the framework of the effective medium theory. The averaged Green's function over different disorder samples can be written as 
\begin{equation}
\overline{G}(\omega)  \equiv \langle \langle \mathbb{G}(\{V_{\bm i}\}, \omega)  \rangle \rangle =  \frac{1}{\omega - h_0 - \overline{\Sigma}}. \label{self}
\end{equation} 
where $\langle \langle \mathcal{O} \rangle \rangle$ means the averaged result of $ \mathcal{O}$ over different disorder samples. Given a sample of disorder $\{V_{\bm i} \}$, the Green's function is denoted as
\begin{equation}
\mathbb{G}(\{V_{\bm i}\}, \omega) \equiv \frac{1}{G_0 ^{-1}- V - {\Sigma}^{\rm U}}. \label{greens}
\end{equation}
Here, $G_0 = 1/ (\omega - h_0 )$ is the free Green's function without disorder and interaction, and $ {\Sigma}^{\rm U}_{\bm{i}\alpha,\bm{j}\gamma} \equiv U\langle \hat{c}_{\bm{i}\alpha}\hat{c}_{\bm{i}\bar{\alpha}}^{\dagger}\rangle \delta_{\bm i \bm j} \delta_{\bar{\alpha}\gamma}+U\langle \hat{c}_{\bm{i}\bar{\alpha}}^{\dagger}\hat{c}_{\bm{i}\bar{\alpha}}\rangle  \delta_{\bm i \bm j}  \delta_{\alpha\gamma} $ is the self-energy within the Hartree-Fock approximation for the weakly interacting case.  The matrix $V$ refers to the disorder potential, with elements $V_{\bm i\bm j} = V_{\bm i} \delta_{\bm i \bm j}$.

The averaged Green's function can be evaluated by expanding the formula\,\eqref{greens} up to the order of $W^2$ and $U$. Then we have 
\begin{equation}
 \langle \langle \mathbb{G}(\{V_{\bm i}\}, \omega)  \rangle \rangle = \left[1 +  {G}_{0}  \langle \langle V \rangle \rangle   +   {G}_{0} \langle \langle   {\Sigma}^{\rm U}  \rangle \rangle  +  {G}_{0}    \langle \langle V  {G}_{0} V \rangle \rangle \right] {G}_{0} (\omega).
\end{equation}
Denoting $\overline{\Sigma}^{\rm U} \equiv \langle \langle   {\Sigma}^{\rm U} \rangle \rangle $ and using the fact that $ \langle \langle V \rangle \rangle =0$ for the random events governed by the uniform distribution in  $[-W, W]$, we get 
\begin{equation}
\overline{G}^{-1} (\omega) ={G}_{0}^{-1} (\omega)  \left[1  -   {G}_{0}\overline{\Sigma}^{\rm U}   -  {G}_{0}    \langle \langle V  {G}_{0} V \rangle \rangle \right].
\end{equation}
According the definition of the self-energy\,\eqref{self} in the effective medium theory, we find
\begin{equation}
\overline{\Sigma}  = \overline{\Sigma}^{\rm U}  +  \langle \langle V  {G}_{0} V \rangle \rangle .
\end{equation}
Since the disorder potential at different positions is independent, we have $ \langle \langle V_{\bm i}   V_{\bm j}  \rangle \rangle =   \delta_{\bm i \bm j} {W^2}/{3} $  for the uniform probability distribution. Finally, we get 
\begin{equation}
 \overline{\Sigma}_{\bm i \bm j }  = \delta_{\bm {i j}} \left\{\overline{\Sigma}^{U}_{\bm i \bm i}  +  {G}_{0,{\bm i \bm  i}} \langle \langle V_{\bm i}^2    \rangle \rangle\right\}. \label{self3}
\end{equation}
Note that the averaged Hartree-Fock self-energy $\overline{\Sigma}^{U}_{\bm i \bm i}$ can be obtained by evaluating the averaged value of the local operators $ \hat{c}_{\bm{i}\alpha}\hat{c}_{\bm{i}\bar{\alpha}}^{\dagger}$  and $ \hat{c}_{\bm{i}\bar{\alpha}}^{\dagger}\hat{c}_{\bm{i}\bar{\alpha}}$. These quantities can be calculated by using the effective Hamiltonian, $H_{\rm{eff}} = h_0 +\overline{\Sigma}$. The Green's function $G_0$ in Eq.\,\eqref{self3} can also be replaced by $G$. The difference of the final results is of higher order. 



\end{widetext}

\end{document}